# Similarity of the Temperature Profile formed by Fluid Flow along a Wall

**David Weyburne**[1]
AFRL/RYDH
2241 Avionics Circle
WPAFB, OH 45433

**ABSTRACT**

A new approach to the study of similarity of temperature profiles is presented. It is applicable for any 2-D fluid flow along an isothermal heated (cooled) wall. The approach is based on a simple concept; the area under a set of scaled temperature profile curves that show similar behavior must be equal. This leads to a new integral-based definition of temperature profile similarity. By taking simple area integrals of the scaled temperature profile and its first derivative, we also obtain a number of new results pertaining to similarity of the temperature profiles. For example, it is shown that if similarity exists, then: 1) the similarity temperature and length scaling parameters are interdependent, 2) the thermal displacement thickness must be a similar length scaling parameter, and 3) the temperature scaling parameter must be proportional to the free stream minus wall temperature values.

[1] David.Weyburne@gmail.com

## 1. INTRODUCTION

Prandtl introduced the concept a boundary layer for fluid flow past a solid over a hundred years ago [1]. The boundary layer concept for flow over a heated flat plate is depicted in Fig. 1. As the flowing fluid impinges on the flat plate a boundary layer is formed along the plate such that the velocity and temperature at the wall gradually transition to the bulk values. Reynolds [2] work on dimensional analysis in the 1800's led to the concept of profile similarity (also called self-similarity). This is the situation in which for certain fluid flows, the downstream velocity or temperature profile along the wall is a simple stretched version of the upstream profile.

For 2-D wall-bounded flows like that depicted in Fig. 1, temperature similarity is defined as the case where two temperature profile curves from different stations along the wall in the flow direction differ only by a scaling parameter in $y$ and a scaling parameter of the temperature profile $T(x,y)$. The temperature profile, in this case, is defined as the temperature taken at all *y* values starting from the wall moving outwards at a fixed *x* value. To put this in formal mathematical terms, we first define the length scaling parameter as $\delta_s$ and the temperature scaling parameter as $T_s$. The temperature profile $T(x_1,y)$ at position $x_1$ along the wall will be similar to the temperature profile $T(x_2,y)$ at position $x_2$ if

$$\frac{T\left(x_1,y/\delta_s\left(x_1\right)\right)}{T_s} = \frac{T\left(x_2,y/\delta_s\left(x_2\right)\right)}{T_s} \quad \text{for all y.} \tag{1}$$

An important observation about the similarity definition is that in the limit of $y \to 0$, the ratio becomes

$$\lim_{y\to 0}\left[\frac{T\left(x,y/\delta_s\left(x\right)\right)}{T_s}\right] = \frac{T_w}{T_s} \quad . \tag{2}$$

Since $T_w$ is a constant (isothermal plate) then similarity requires that $T_s$ must also be a constant. This means that only the length scaling parameter $\delta_s$ may vary with the flow direction (x-direction).

Traditionally, similarity of the temperature profile involved finding solutions to the flow governing equations. Schlighting [3], for example, reviews some of the laminar flow temperature profile similarity solutions obtained by this approach. More recently, Wang, Castillo, and Araya [4] and Saha, *et al*. [5] among others, have discussed this approach for the forced turbulent boundary layer flow case. Recently, Weyburne [6] introduced a new approach for studying similarity of the velocity profile for 2-D flow along a wall. In the work herein, we apply this same equal area, integral approach to study the temperature profile similarity formed by 2-D flow over a heated (cooled) isothermal wall. The approach is based directly on Eq. 1. We start by manipulating the similarity definition equation and then integrating the result. For example, in the Section below we differentiate both sides of Eq. 1 with respect to the scaled y-coordinate and then integrate the result from the wall to a position deep into the freestream. This area integral is easily manipulated and can be solved explicitly as a ratio of parameters. Similarity necessarily requires that the area under the curves must be equal which then imposes an equivalence condition on the parameter ratios. Since the results are derived directly from the definition of similar curves given by Eq. 1, we, therefore, do not need to know how the curves were generated. Hence, the flow governing equations do not have to be invoked. Furthermore, the new results apply to similar profile curves whether they are laminar, transitional, or turbulent fluid flows as long as the temperature is taken as the Reynolds averaged temperature.

Experimental laminar and turbulent datasets are examined in order to verify the new results. For the laminar flow case, experimental wind tunnel results are not available so we turn to the theoretical solutions offered by Pohlhausen [7]. For the turbulent boundary layer flow, we use the tabulated data from Blackwell, Kays, and Moffat [8] and R. du Puits, C. Resagk and A. Thess [9].

## 2. Similarity of the Temperature Profile

### 2.1 The First Derivative Profile

Now we turn to the task of considering what can be learned from this equal area approach to similarity for 2-D flow along a heated (cooled) isothermal wall. In the analysis below, the only assumptions that are necessary as to the functional form of the temperature profile is that the form is consistent with a physically realizable form, *i.e*. no discontinuities or singularities. The development presented herein is based on a simple concept; for similarity, the area under the scaled temperature profile curves plotted versus the scaled *y*-coordinate must be equal at each station along the wall.

Rather than starting directly with the consideration of the temperature profile, we will first consider the implications of similarity on the first derivative profile curves since this result is a simple illustration of the new approach and its result is required later on. We start with the formal definition of similarity given by Eq. 1. If similarity is present in a set of temperature profiles then it is self-evident that the properly scaled first derivative profile curves (derivative with respect to the scaled *y*-coordinate) must also be similar. It is also self-evident that the area under the scaled first derivative profiles plotted against the scaled *y*-coordinate must be equal for similarity.

In mathematical terms, the area under the scaled first derivative profile curve is given by

$$a(x) \;=\; \int_0^{h/\delta_s} d\left\{\frac{y}{\delta_s}\right\} \frac{d\left\{T\left(x, y/\delta_s\right)/T_s\right\}}{d\left\{\dfrac{y}{\delta_s}\right\}} \quad , \tag{3}$$

where $a(x)$ is in general a non-zero numerical constant. For clarity, $h(x)$ and $\delta_s(x)$ have been shortened to $h$ and $\delta_s$. For equal areas at each measurement station, it is required that the integral limit condition $h(x_1)/\delta_s(x_1) = h(x_2)/\delta_s(x_2)$ must hold. However, this requirement is easily satisfied since we can freely choose $h(x_1)$ and $h(x_2)$ as long as they are both located deep into the free stream above the boundary layer edge. Using the boundary conditions $T(x,0) = T_w$ and $T(x,h) = T_\infty$ (where $T_\infty$ is the free-stream temperature above the boundary layer edge and $T_w$ is the wall temperature) and a simple variable switch $\left(d\{y/\delta_s\} \Rightarrow (1/\delta_s)dy\right)$, Eq. 3 can be shown to reduce to

$$a(x) \quad = \quad \int_0^{h/\delta_s} d\left\{\frac{y}{\delta_s}\right\} \frac{d\left\{T\left(x,y/\delta_s\right)/T_s\right\}}{d\left\{\frac{y}{\delta_s}\right\}} \quad , \tag{4}$$

$$a(x) \quad = \quad \int_0^{h} dy \; \frac{d\left\{T(x,y)/T_s\right\}}{dy}$$

$$a(x) \quad = \quad \frac{1}{T_s}\int_0^{h} dy \; \frac{d\,T(x,y)}{dy}$$

$$a(x) \quad = \quad \frac{1}{T_s}\left[T(x,y)\right]_{y=h,0}$$

$$a(x) \quad = \quad \frac{T_\infty - T_w}{T_s} \quad .$$

Similarity requires that $a(x_1) = a(x_2) = \text{constant}$. The implications of this similarity requirement are obvious; the similarity scaling parameter $T_s$ must be proportional to the temperature difference $T_\infty - T_w$ .

## 2.2 Similarity of the Defect Profile

Before we turn our attention to the scaled temperature profiles, we first need to point out that in the limit of *y* going deep into the free stream, Eq. 1 requires

$$\lim_{y\to h(x_1)}\left[\frac{T\left(x_1,y/\delta_s\left(x_1\right)\right)}{T_s}\right] \quad = \quad \lim_{y\to h(x_2)}\left[\frac{T\left(x_2,y/\delta_s\left(x_2\right)\right)}{T_s}\right] \quad = \quad \frac{T_\infty}{T_s} \quad = \quad \text{constant} \quad . \tag{5}$$

With this identity, we are now in a position to turn our attention to the scaled temperature profiles. Starting with the formal definition of similarity given by Eq. 1 then it is self-evident that for the profiles to be similar, the area under these scaled temperature profiles plotted versus the scaled *y*-coordinate must be equal. The area under the scaled temperature profile is not easily manipulated as is done above. However, we can use the results from the last Section to advantage. If we add or subtract a constant to both sides of Eq. 1 and then integrate, the equivalence condition still holds. The Eq. 5 implication is that $T_\infty / T_s$ must be a constant if similarity is present. Subtracting this constant value from both sides of Eq. 1 and integrating, the area, in mathematical terms, is given by

$$c_0(x) \quad = \quad \int_0^{h/\delta_s} d\left\{\frac{y}{\delta_s}\right\} \frac{T\left(x,y/\delta_s\right) - T_\infty}{T_s} \quad , \tag{6}$$

where $c_0(x)$ is in general a nonzero numerical constant. Using a simple variable switch and simple algebra, Eq. 6 can be shown to reduce to

$$c_0(x) = \int_0^{h/\delta_s} d\left\{\frac{y}{\delta_s}\right\} \frac{T(x, y/\delta_s) - T_\infty}{T_s} \quad , \tag{7}$$

$$c_0(x) = \frac{1}{\delta_s}\int_0^{h} dy \; \frac{T(x,y) - T_\infty}{T_s}$$

$$c_0(x) = \frac{T_w - T_\infty}{\delta_s \, T_s}\int_0^{h} dy \; \frac{T(x,y) - T_\infty}{T_w - T_\infty}$$

$$c_0(x) = \frac{\delta_T^* \, (T_w - T_\infty)}{\delta_s \, T_s}$$

where the $\delta_T^*$ is the thermal displacement thickness given by

$$\delta_T^*(x) = \int_0^{h} dy \; \frac{T(x,y) - T_\infty}{T_w - T_\infty} \quad , \tag{8}$$

and where $h$ is located deep in the free stream. Eq. 9 is the direct analog of the velocity profile displacement thickness and is in fact the area under the scaled defect profile.

Eq. 7 is an exact result whether the profiles are similar or not. Similarity requires that $c_0(x_1) = c_0(x_2) = \text{constant}$. The importance of Eq. 7 in regards to similar profiles is that it means that the thickness scaling factor $\delta_s$ and temperature scaling factor $T_s$ are not independent for 2-D wall-bounded similarity flows.

Using the result given by Eq. 7 combined with the result given by Eq. 4, then it is evident that $c_0(x)$ reduces to

$$c_0(x) = \frac{\delta_T^*(x)}{\delta_s(x)} \quad . \tag{9}$$

Similarity requires that $c_0(x_1) = c_0(x_2) = \text{constant}$. This equivalence condition on Eq. 9 means that if similarity is present in a set of temperature profiles, then the thermal displacement thickness must be a length scale that results in similarity.

## 2.3 Integral Definition of Similarity

Having equal $c_0(x)$ values (Eq. 6) at different stations along the flow is a necessary but not a sufficient condition for similarity of a set of profiles. If the scaled temperature profiles are similar, then it is self-evident that the scaled temperature profiles multiplied by the scaled $y$-coordinate raised to the $n$th power must also be similar. In mathematical terms, the area under the scaled temperature profiles multiplied by the scaled $y$-coordinate raised to the $n$th power is given by

$$c_n(x) = \int_0^{h/\delta_s} d\left\{\frac{y}{\delta_s}\right\}\left(\frac{y}{\delta_s}\right)^n \frac{T(x, y/\delta_s) - T_\infty}{T_s} \quad , \tag{10}$$

where $c_n$ are, in general, non-zero numerical constants. Mathematically, it is self-evident that a sufficient condition for similarity of a set of profiles is that

$$c_n(x_1) \;=\; c_n(x_2) \qquad \text{for } n=0,1,2,3,\ldots,\infty \;, \tag{11}$$

so long as $h$ is chosen appropriately and is located deep in the free stream.

## 3. Experimental Verification

### 3.1 Laminar Flow

The experimental verification section is not intended to be an exhaustive comparison to various experimental results. The above theoretical results are exact and unimpeachable. So instead, this section is simply intended to demonstrate that there is experimental support for the theory. For the laminar boundary layer, we were not able to find experimental wind tunnel data to test the new similarity scaling parameters so instead, we turned to the similarity solutions to the flow governing equations. For the laminar flow case, the Pohlhausen [7] theoretical solution method for laminar flow on a heated isothermal flat wall is an appropriate first step. In Fig. 2a the results for the Pohlhausen-based approach to calculating thermal profiles for a range of Prandtl numbers is presented. This figure is a re-creation of Fig. 12.9 from Schlichting [3] using a simple FORTRAN program to generate the solutions. In this figure, the reduced temperature defined as

$$\theta(\eta) \;=\; \frac{T(\eta)-T_\infty}{T_w - T_\infty} \;, \tag{13}$$

is plotted against the reduced Blasius height given by

$$\eta \;=\; y\sqrt{\frac{u_\infty}{\nu_\infty x}} \;, \tag{14}$$

where $\nu_\infty$ is the kinematic viscosity.

Now consider Fig. 2b. Although not readily apparent, all nine curves have collapsed on top of one another. It is the same data plotted in Fig. 2a but using the similarity length scale parameter $\delta_T^*$; the thermal displacement thickness (Eq. 8). Note that the Phohlhausen solution method using the reduced temperature $\theta$ has already fixed the temperature similarity scaling parameter as $T_s \propto (T_\infty - T_w)$.

### 3.2 Turbulent Flow

For the turbulent boundary layer, the number of available thermal turbulent boundary layer datasets is limited. One of the most extensive available sets is from Blackwell, Kays, and Moffat [8]. In that study, an experimental investigation of the heat transfer behavior of the near-equilibrium transpired turbulent boundary layer with adverse pressure gradient was carried out. In particular, adverse pressure gradients of the form $u_e(x) \sim x^m$, m = 0, -0.15, and -0.2, were considered along with a variety of transpired conditions ($u_e(x)$ is the velocity at the boundary layer edge). The data which closely matches the conditions depicted in Fig. 1 is the non-transpired, near-equilibrium 091871, 120471, and 110871 datasets.

The data presented in Fig. 3a represents the last six out nine curves from the m = 0 dataset, the last seven out of nine curves from the m = -0.15 dataset, and the last three of nine curves from the m = -0.2 dataset. In Fig. 3b the same data is presented as defect profiles. It is evident that both the scaled temperature profiles and the scaled defect profiles display similarity-like behavior. These two plots indicate that neither the scaled temperature nor the scaled defect profiles collapse to some universal function of the scaled height as is sometimes claimed in the literature (see for example [4]).

Finally, R. du Puits, C. Resagk and A. Thess [9] have made highly resolved temperature measurements in turbulent Rayleigh-Bénard convection in air at a Prandtl number *Pr* = 0.7. Although not technically forced flow, the resulting gravity induced recirculation closely match the forced flow conditions at certain locations. In Fig. 4, we plot the reduced temperature versus the scaled height for one of their datasets that matches this condition. We note that the reduced temperature data, $\theta^*$, was corrected for the fact that the as reported data never asymptote to the bulk temperature. This correction was done by simply rescaling the reported $\theta$ data so that it spans 0 to 1. We note that R. du Puits, C. Resagk and A. Thess [9] did, in fact, report that the scaled temperature showed similar behavior when scaled by the thermal displacement thickness, $\delta_T^*$. The theoretical results herein confirm their experimental observation.

## 4. Discussion

An important point about the above derivations is that although the results are not presented formally, the results above are mathematically rigorous and can be easily substantiated in the form of mathematical proofs. The above results apply to any 2-D fluid flowing along a heated (cooled) isothermal wall including all laminar flows, transitional flows, or turbulent fluid flows as long as the temperature is taken as the Reynolds averaged temperature.

The significance of the equivalence condition on Eq. 4 is that it indicates that if similarity exists in a set of temperature profiles then the appropriate temperature scaling parameter for 2-D wall-bounded flow along an isothermal wall is $T_\infty - T_w$. In addition, the equivalence condition on Eq. 9 indicates that if similarity exists in a set of temperature profiles then the thermal displacement thickness $\delta_T^*$ must be a similarity length scale factor. To our knowledge, this is the first time the thermal displacement thickness has been theoretically identified as a similarity length scaling parameter in the literature.

The experimental verification presented above is intended to simply demonstrate that the new similarity scalings do work for certain data sets. The results for the laminar flow case are especially significant in that Pholhausen's [7] laminar flow work appears in every major textbook that deals with heat and fluid flow. The collapse of the figures to one curve (Fig. 2b) using $\delta_T^*$ as the stretching parameter is dramatic. This the first time the thermal displacement $\delta_T^*$ has been even considered as a similarity scaling parameter to our knowledge. The results herein prove that that must be the case for similarity to be present in any 2-D wall-bounded flow.

Although these results apply to any flow showing similar behavior, the case of the turbulent boundary layer is interesting. Prior to the work herein, the theoretical approach to similarity for turbulent boundary flows has been hampered by the fact that the flow governing equations do not have a closed form for turbulent flows. Hence, closed form mathematical solutions do not exist as is the case for laminar flows. Therefore, what is usually done is to make simple educated guesses for what might be the proper similarity scaling and then compare those guesses to experimental results. This is the path taken by Wang, Castillo, and Araya [4] and Saha, *et al*. [5] among others. We have not attempted to compare the results herein to these previous guesses precisely because the results herein are mathematically derived, not guessed.

Mathematically, Eq. 1 is the traditional definition of temperature profile similarity but a mathematically equivalent definition is to require the condition given by Eq. 11 exists for all $n$. For use on experimental datasets, this latter method would seem to have the untenable requirement of calculating a very large (infinite) set of integrals. However, for flow similarity of temperature profiles taken at various stations along the wall in the flow direction, the profile curves are not arbitrary and, in general, are not changing rapidly. Hence it may only be necessary to ensure that the first couple of $c_0(x_1)=c_0(x_2)$ values are constant as opposed to the infinite set. In any case, this approach to similarity has an advantage from an experimental standpoint since the equal area test method would allow for statistical testing for similarity by comparing $c_n(x)$ values at various stations along the wall. From a practical standpoint, this method is superior to the use of Eq. 1 since to use Eq. 1 the experimentalist needs to insure the measured temperature at each $y/\delta_s(x)$ value is equal at both measurement stations. This is a very difficult task since $\delta_s(x)$ is changing with $x$ and, in general, its value is not known *a priori*. As a result of the $y/\delta_s(x)$ issue, the usual imprecise method the flow community presently uses to judge whether a set of experimental temperature profiles are similar is to plot the profiles and use the subjective "chi-by-eye" method to judge the success or failure (do they plot on top of one another?). In contrast, the equal area test method would allow for statistical testing by performing simple numerical integrals of the profile data.

## 5. Conclusion

A new approach for studying similarity of the temperature profile is outlined. It starts from the equation used to define similarity of the temperature profile. This method was used to discover fundamentally new results for the similarity of 2-D flow along a heated (cooled) isothermal wall. It was shown that if similarity exists, then the similarity temperature and length scaling constants cannot be independent. Furthermore, it was shown that if similarity exists, the thermal displacement thickness $\delta_T^*$ must be a length scaling variable and the temperature difference $T_\infty - T_w$ must be a temperature scaling parameter.

**ACKNOWLEDMENTS**

This work was supported by AFOSR, PM Dr. Gernot Pomrenke, and the Air Force Research Laboratory. The author thanks the various experimentalists for making their datasets available for inclusion in this manuscript.

**Figures**

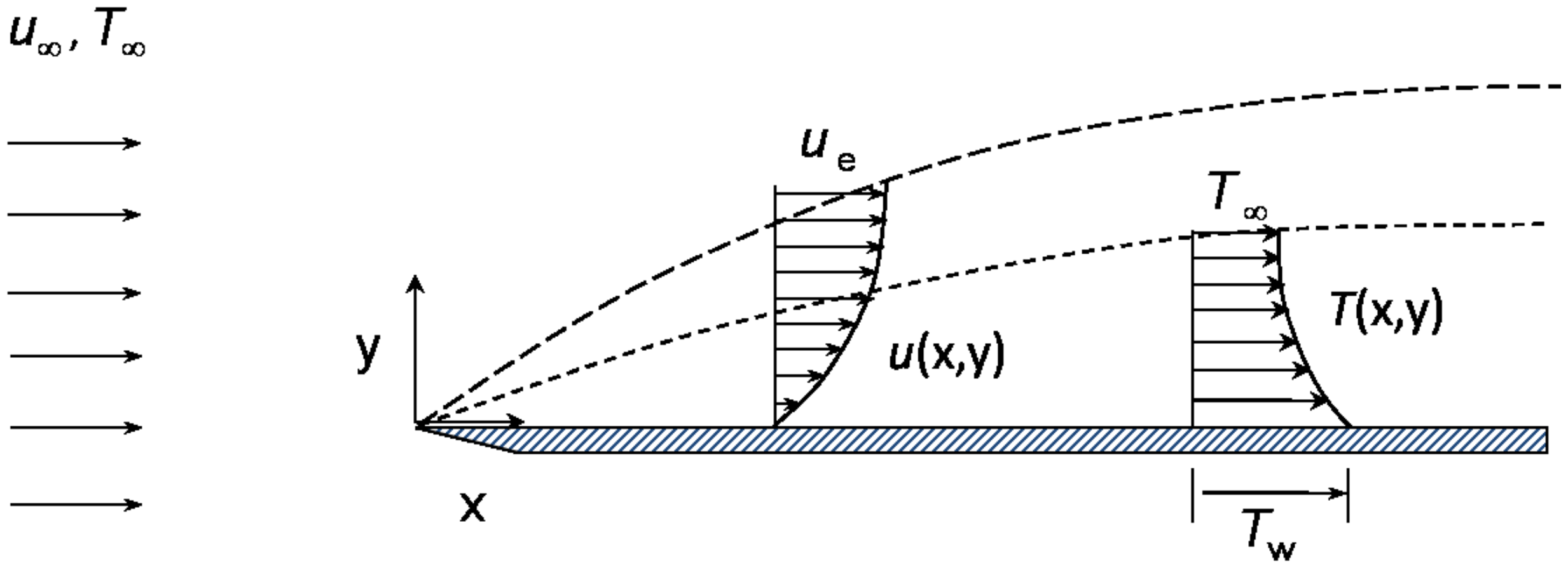

Fig. 1: A schematic diagram showing the flat plate 2-D flow geometry and variables.

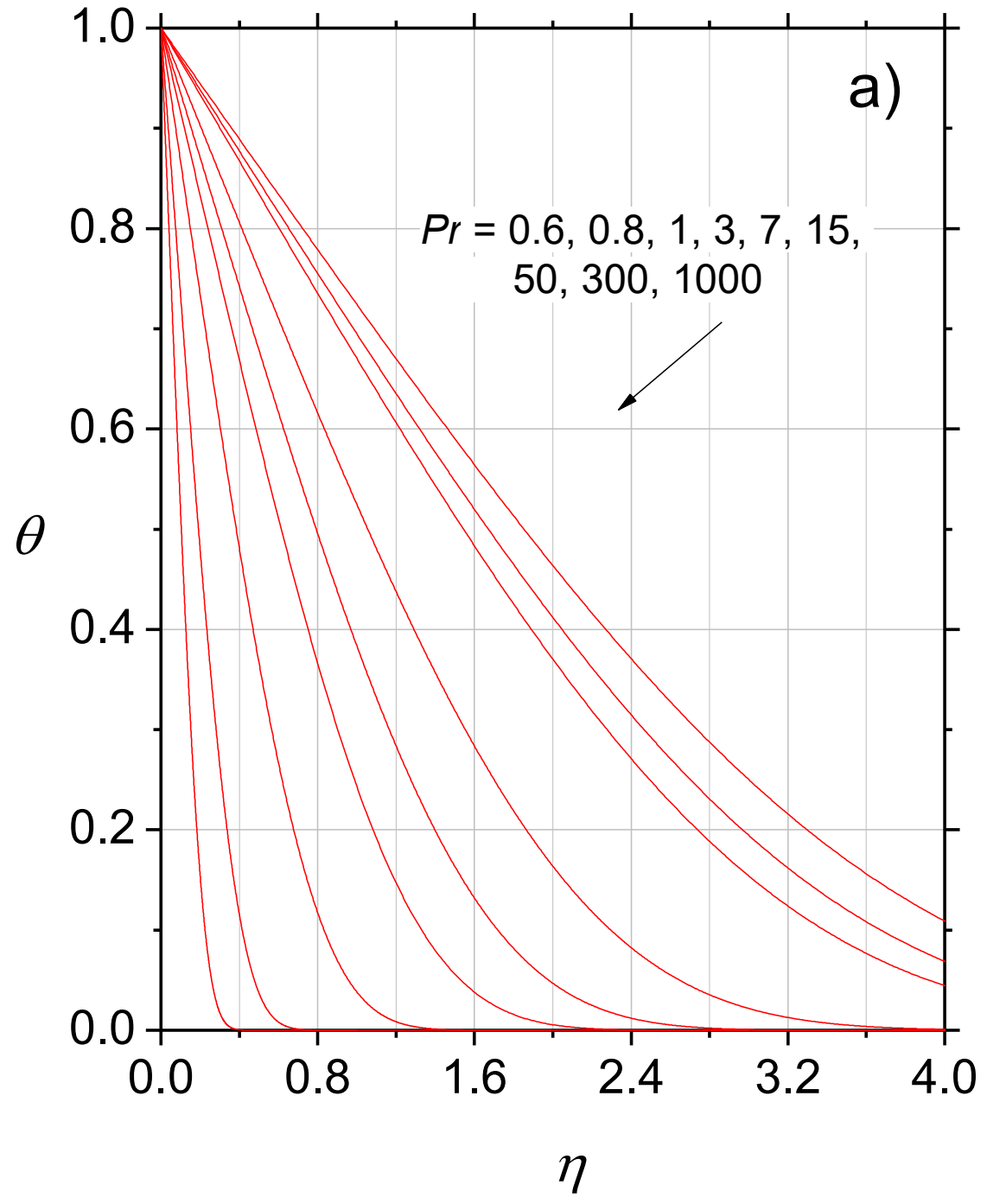

Fig. 2a: Scaled temperature profiles for laminar flow over a plate for a range of *Pr* numbers (after Schlichting [3]).

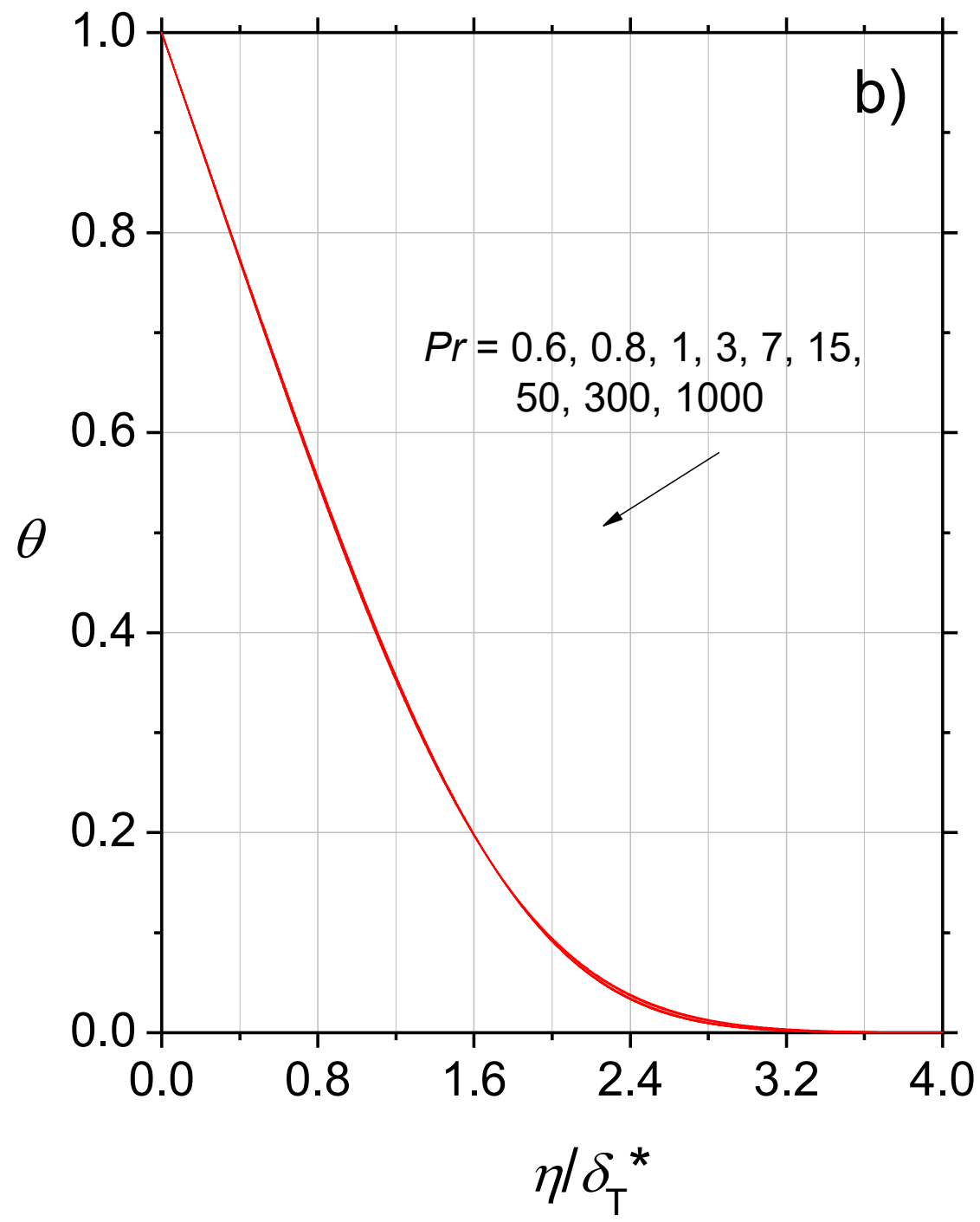

Fig. 2b: The same temperature data from Fig. 2a but with the new length scale. All nine curves are collapsed onto one another.

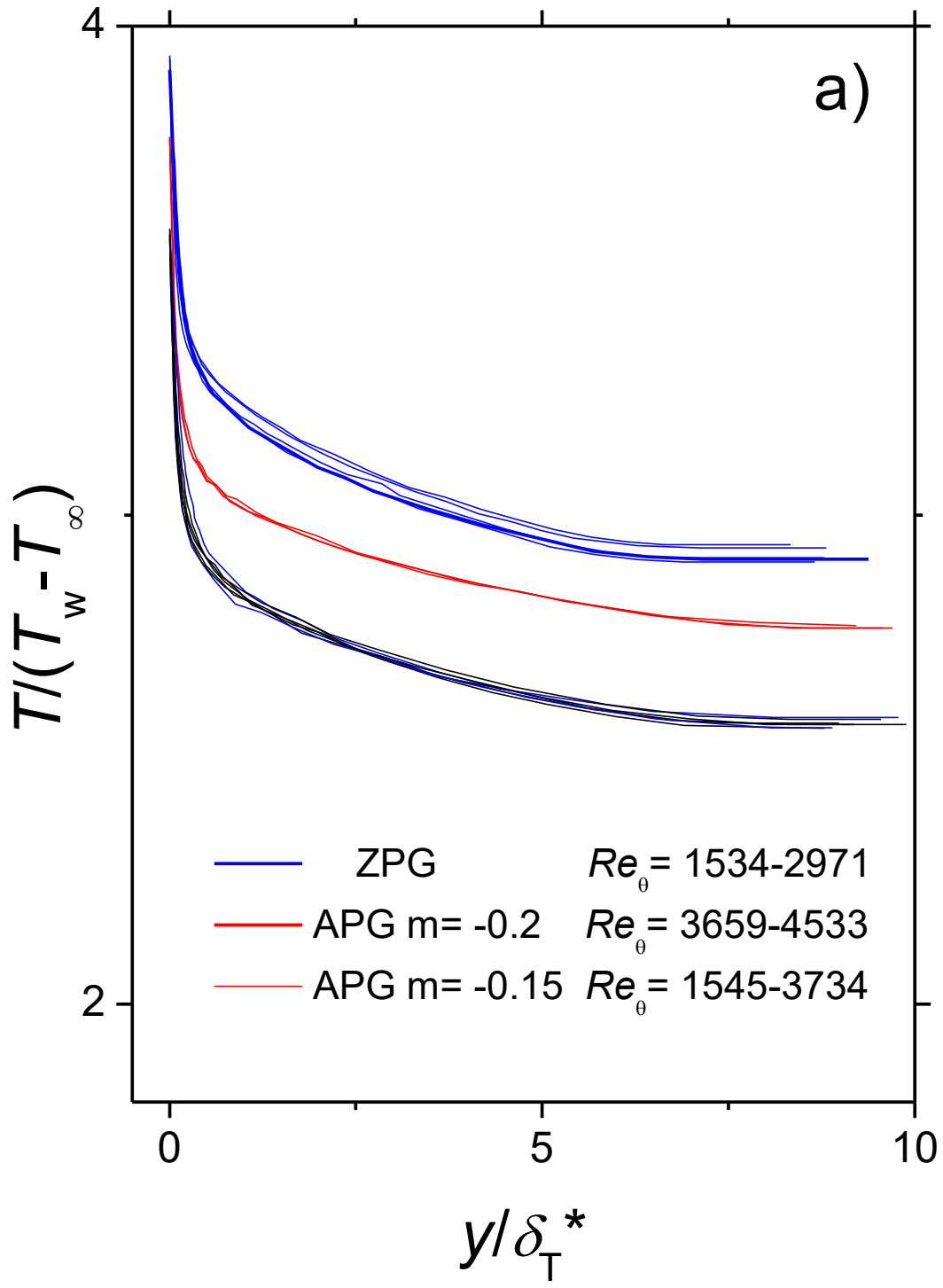

Fig. 3a: Scaled temperature profiles for Blackwell, Kays, and Moffat [5] turbulent boundary layer flow.

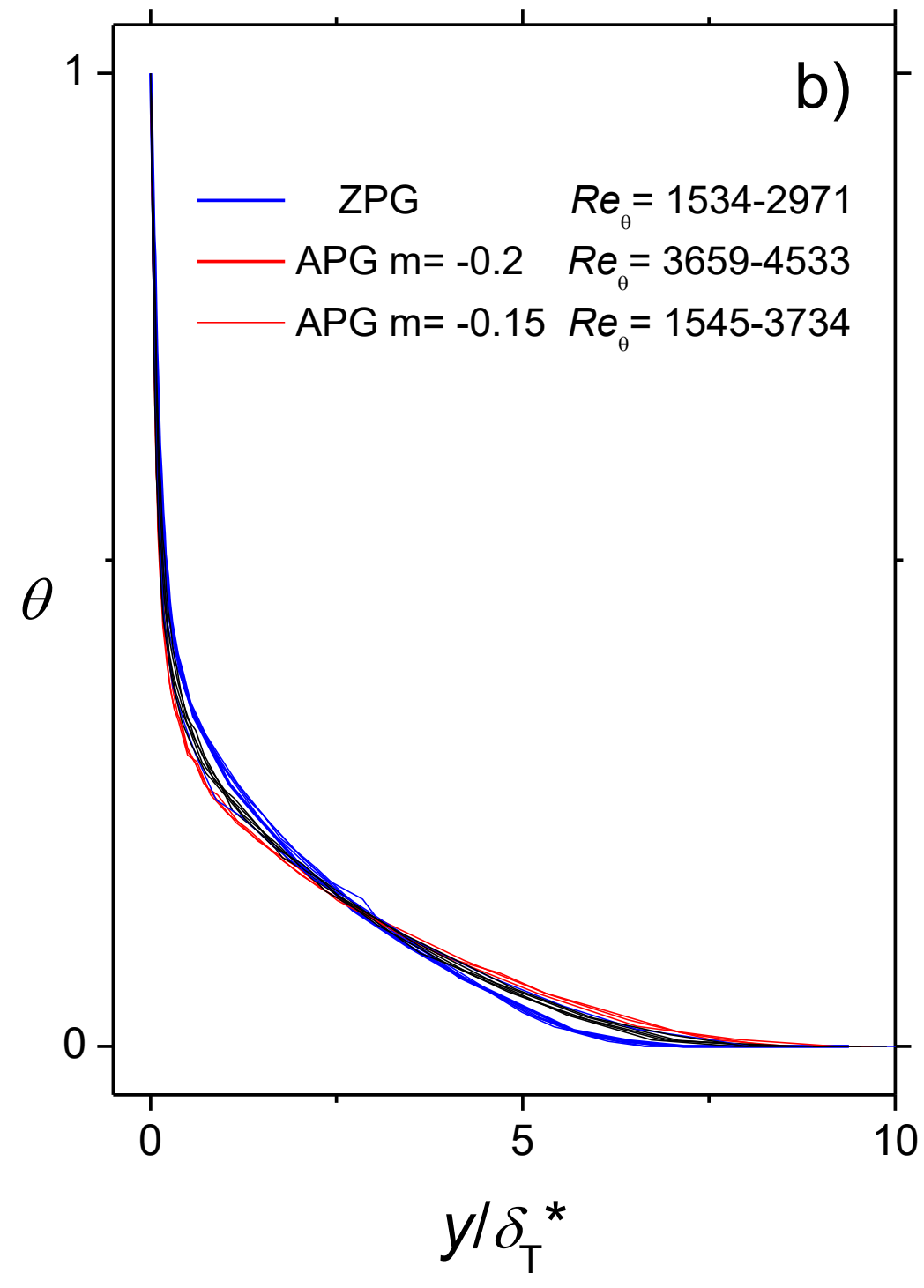

Fig. 3b: The same scaled temperature profile data from Fig. 3a but scaled as defect profiles.

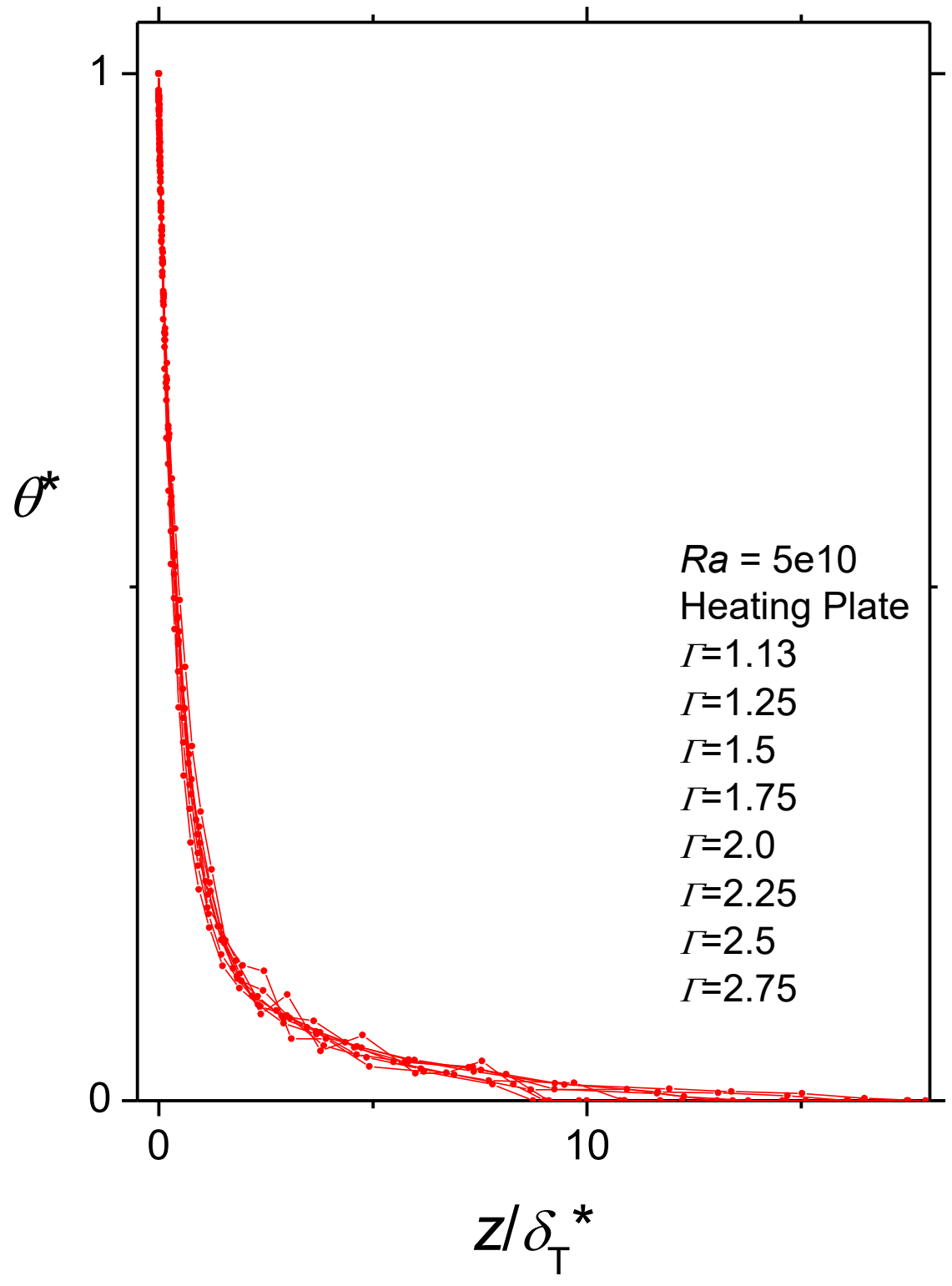

Fig. 4: Eight scaled defect temperature profiles from R. du Puits, C. Resagk and A. Thess [9] Rayleigh-Bénard turbulent boundary layer flows.